\documentclass[traditabstract]{aa}
\usepackage{graphicx}
\usepackage{txfonts}
\newcommand \htwo  {\mbox{H$_2$ }}

\newcommand \degree {\mbox{$^\circ$}}
\newcommand \msun {\mbox{$\mathcal{M}_{\odot}$}}

\newcommand \kms {\mbox{km s$^{-1}$}}

\newcommand \twelvecoonezero {\mbox{$^{12}$CO(1--0) }}
\newcommand \twelvecotwoone  {\mbox{$^{12}$CO(2--1) }}

\newcommand \hcn {\mbox{HCN(1--0) }}

\newcommand \halfa {$\mbox{H}\alpha$}
\newcommand \ergcms {$\mbox{ergs cm}^{-2}\mbox{s}^{-1}$}
\newcommand \msunyr {\mbox{$\mathcal{M}_{\odot}\mbox{yr}^{-1}$}}
\newcommand \tausf  {\mbox{$\tau_{\mbox{\small SF}}$}}

\begin{document}
   \title{HCN(1-0) enhancement in  the bar of NGC~2903}


   \author{S. Leon \inst{1},
        S. Jeyakumar \inst{2}
          D. P\'erez-Ram\'{\i}rez, \inst{3}
          L. Verdes-Montenegro, \inst{4}
           S.W. Lee, \inst{5}
	   B. Oca\~na Flaquer \inst{1}
          }

	\authorrunning{Leon et al.}

   \institute{Instituto de Radioastronom\'{i}a Milim\'etrica (IRAM), Avenida Divina Pastora 7, N\'{u}cleo Central, 18012 Granada, Spain
           \and
             Departamento de Astronomía, Universidad de Guanajuato, AP 144, Guanajuato CP 36000, Mexico.
	   \and
	       University of Ja\'en, Ja\'en, Spain 
           \and	
               Instituto de Astrof\'{\i}sica de Andaluc\'{\i}a-CSIC, Granada, Spain 
           \and
               University of Toronto, Astronomy Department, Toronto, Canada 
             }

   \date{Received XX; accepted XX}

 
  \abstract
{
We have mapped the \hcn emission from two spiral galaxies, 
NGC~2903 and NGC~3504 to study the gas properties in the bars. 
The \hcn emission is detected in the center 
and along the bar of NGC~2903. The line ratio \hcn/\twelvecoonezero 
ranges from 0.07 to 0.12 with the lowest value in the center. 
The enhancement of \hcn emission along the bar indicates a higher fraction of 
dense molecular gas in the bar than at the center. 
The mass of dense molecular gas in the center ($2.2\times 10^7$ \msun) is  about 6 times lower than
that in the bar ($1.2\times 10^8$ \msun). The total star formation rate (SFR) is estimated to be 1.4 \msunyr,
where the SFR at the center is  1.9 times higher than that in the bar.
The time scale of consumption of the dense molecular gas in the center 
is about $\sim 3\times 10^7$ yr which is much shorter than that in the bar of 
about 2 to 10$\times 10^8$ yr.  The dynamical time scale of inflow
of the gas from the bar to the center is shorter than the consumption 
time scale in the bar, which suggests that the star formation (SF) activity at the center is not 
deprived of fuel.  In the bar, the fraction of dense molecular gas mass
relative to the total molecular gas mass is twice as high along the leading edge 
than along the central axis of the bar. 
The \hcn emission has a large velocity dispersion in the bar,
which can be attributed  partially to the streaming motions  indicative of shocks
along the bar. 
In NGC~3504, the \hcn emission is detected only at the center. The
fraction of dense molecular gas mass in the center is about 15\%.
Comparison of the SFR with the predictions from numerical 
simulations suggest that NGC~2903 harbors a young type B bar with a strong 
inflow of gas toward the center whereas NGC~3504 has an older bar and 
has already passed the phase of inflow of gas toward the center. 
}{}{}{}{}

   \maketitle
%

\section{Introduction}

The molecular interstellar medium in galaxies has been extensively
studied through the rotational transitions of CO at millimeter wavelengths. 
These transitions are
good tracers of the molecular gas mass  and represent
the general distribution of molecular hydrogen (e.g. Young \& Scoville
1982, Young \& Devereux 1991).
High-density gas tracer molecules, like HCN, add  relevant information to this view
concerning the dense gas ($n_{H_2} > 10^4 cm^{-3}$).
Observational evidence 
(Nguyen et al. 1992, Reynaud \& Downes 1997, Kohno et al. 1999a)
has suggested a close relationship between dense molecular gas and
massive star formation in the centers of galaxies. In the
center of the starburst galaxies, the HCN line emission is tightly correlated with the
radio-continuum emission (see e.g. in NGC~1530, Reynaud \& Downes 1997)
and the total HCN luminosity correlates with the
far-infrared (FIR) luminosity (Solomon et al. 1992, Gao \& Solomon 2004), although contradictory 
results have been found, e.g. by Aalto et al. (1995).
Since only very few galaxies have been mapped in HCN line
emission at large scales, this correlation remains a source of
speculation, especially  for milder star formation (SF),  typically 2 \msunyr or less.

The SF in the centers of galaxies and in the spiral arms has been extensively 
studied.  Nevertheless the processes involved in the SF activity in the bar 
itself are  still not well  understood. 
Hydrodynamical and N-body simulations show inflow of gas
along the leading edge after  a shock,  losing
angular momentum (Athanassoula 1992). Sticky particle simulations of the gas in barred galaxies
(Combes \& Gerin 1985) produce the same configuration of enhanced molecular gas along the leading
edge  due to crossing/crowding of orbits. 
The role of gas orbits in a barred potential on the star formation activity is still
poorly understood. 

Various observational studies using CO emission 
have found that molecular gas is located mainly along the leading edge of the bars 
(e.g Handa et al.  1990, Reynaud \& Downes 1997, Downes et al. 1996, 
Sheth et al. 2002). 
However a contribution from large scale diffuse, unbound gas in the
bar of the galaxy NGC~7479 has been suggested by H\"uttemeister et al. (2000). 
 In the bar of NGC~1530, Reynaud \& Downes (1998) have  detected large velocity gradients due to 
velocity jumps between the upstream regions and the shock along the leading edge. These shocks would inhibit
the SF by destroying the giant molecular clouds because of a large shear.


The CO line is a tracer of molecular gas at low densities,
whereas star formation occurs in very dense molecular clouds 
(Solomon et al. 1992, Paglione et al. 1995, Hatchell et al. 1998) which can be 
traced by the HCN transitions. To study the formation of dense molecular  gas  and its relation
to  SF, the bars offer an unique dynamical system  to analyse the effects of a strong density wave
on the molecular gas and SF. Moreover, numerical simulations and H$\alpha$ observations (Martin \& Friedli 1997, Verley et al. 2007) 
have shown a tight  correlation between the SF along the bars and the age of the bars, which suggests that  a similar
correlation can be expected for the \hcn emission in the bar 
 because of the tight relationship between the SF and the dense molecular gas.

In this paper we present the result of the  HCN observations at the IRAM-30m telescope, 
along the bars of two spiral galaxies which have been chosen for their previous detection of 
HCN in the center. In  Sect. 2, the IRAM observations are described. 
The main results of the NGC~3504
and NGC~2903 observations are presented in Sect. 3 and 4 respectively. 
The SF activity is discussed  in Sect. 5 and a 
discussion about the molecular gas properties in the bar  is presented in Sect. 6.

\section{Observations}

\begin{figure*}
\resizebox{16cm}{8.5cm}{\includegraphics{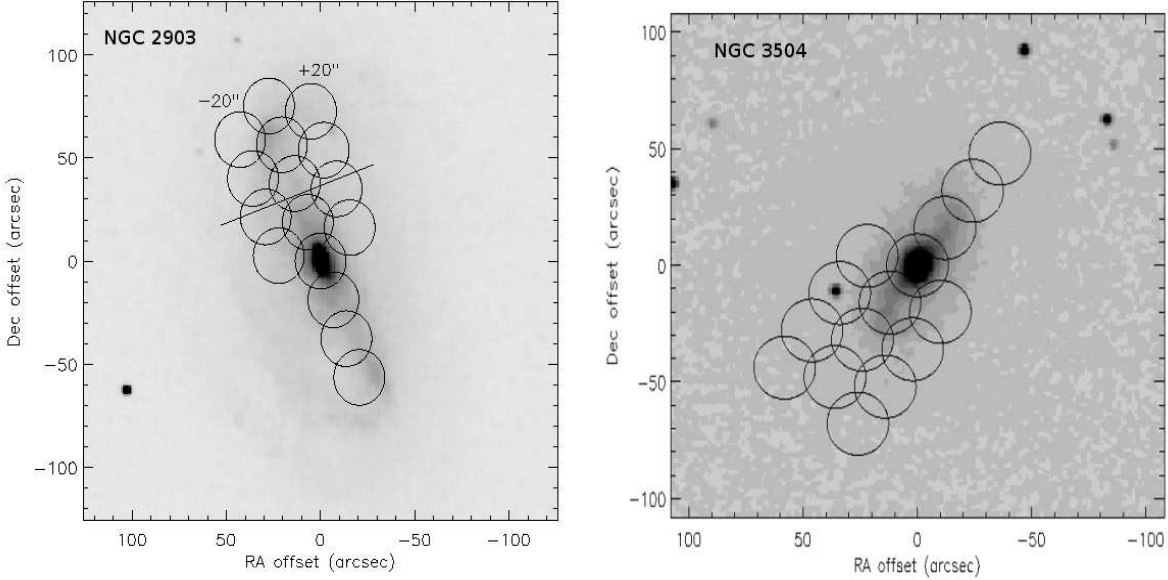}}
\caption{
The positions in NGC~2903  and NGC~3504 of HCN pointing are shown as circles of the size of the beam (27\arcsec) overlaid
on the 2MASS J band image.
The offsets -20\arcsec\ and +20\arcsec in NGC~2903 indicate the leading and trailing edge of the bar. The solid line indicates the cut across which the width of the bar is 
estimated (see Fig. \ref{fig_n2903_cut_bar})}
\label{fig_n2903_2Mass_hcnpos}
\end{figure*}

\begin{figure}
\resizebox{9cm}{8.5cm}{\includegraphics{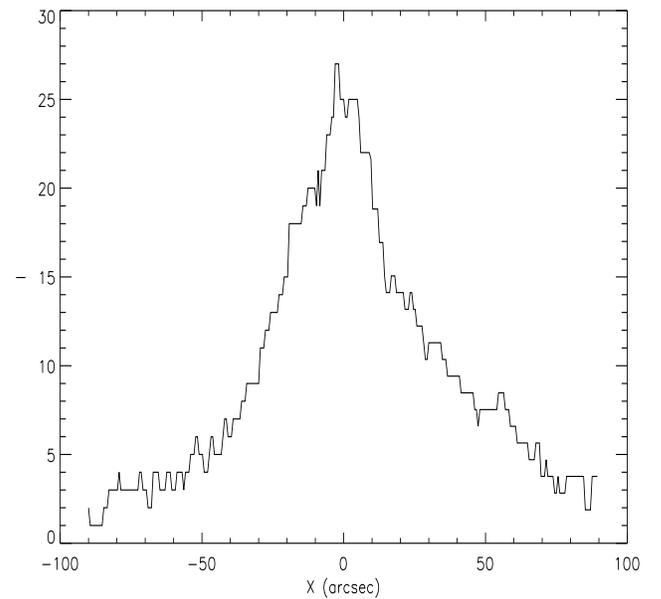}}
\caption{J-band intensity from 2MASS data (arbitrary units)  across the 
cut perpendicular to the bar PA in NGC~2903, at a distance of 30\arcsec 
from the center. The FWHM is estimated to be $48\arcsec$.}
\label{fig_n2903_cut_bar}
\end{figure}

During May and August 2002, we observed 
two barred galaxies, namely NGC~2903 and NGC~3504, using the IRAM-30m telescope.
NGC~2903 and NGC~3504 were observed in good weather 
conditions. The basic parameters of the NGC~2903 and NGC~3504 are given in
the Table~\ref{tab_n2903_n3504}.
The main goal was to map the \hcn emission along the bar of each galaxy. Given the 
beam size at 88 GHz (27\arcsec) and the width of the bar ($\sim$ 48\arcsec, see below) 
for both galaxies, 
the maps were obtained with a sampling of 20\arcsec\ along  the bar and 
$\pm 20$\arcsec\ perpendicular to the bar axis to sample the leading/trailing 
edge as well.  The \twelvecotwoone line was observed in  
parallel, with a spatial resolution
of 10\arcsec\ but  spatially undersampled with the  20\arcsec\ sampling and they are not presented
in this paper, except for the central point in Fig.~\ref{fig_n3504_hcn}.
NGC 3504 was mapped along the bar with a larger coverture on the Southeast side  up to a 
radius of 50\arcsec\ in the bar frame. 
 NGC~2903 was mapped  in \hcn along the bar 
(PA=20\degree) and  along both sides of the bar ($\pm$20\arcsec in the bar frame).
The individual target positions are shown in Fig. \ref{fig_n2903_2Mass_hcnpos}.
To measure  the \hcn on the trailing/leading edge of the bar we  mapped 
from 10\arcsec\ to 70\arcsec\ (2.9 kpc) on the northern side of the bar,
with an offset of $\pm20\arcsec$ perpendicular to the bar axis and a sampling
of 20\arcsec.
The FWHM of the bar structure, estimated using the  J-band image of the Two Micron All Sky Survey, measured
at 30\arcsec from the center after a rotation by -20\degree, is about 48\arcsec
(see Fig. \ref{fig_n2903_cut_bar}). 
For this size of the bar, the pointings at $\pm$20\arcsec\ offsets along the bar 
cover the trailing and leading edge of the bar respectively.

The spectrometer used for the observations 
of the \hcn  line  was the filterbank (512$\times$1MHz channels) and for the \twelvecotwoone 
line  an autocorrelator (409$\times$1.25 MHz resolution) was used.  For the \hcn observations
the typical rms noise temperature is about 4 mK ($T_a^*$ scale) at a velocity resolution 
of 3.4 \kms. After subtracting a  polynomial baseline of order at most 1, the spectra 
were smoothed by the Hanning convolution to obtain a velocity resolution 
of 27 \kms\  with an rms temperature of  1 mK and 5 mK for the observations of NGC 2903 and NGC 3504 
respectively.
Pointing observations on a strong radio source close to the galaxies were  performed every 90 minutes 
with good accuracy, and the expected rms error on the pointing is  
$\sim 3\arcsec$ which is about 10 \% of the beam size at the frequency of 
the \hcn transition. 

\begin{table}
\caption{The basic parameters of the galaxies.}
\label{tab_n2903_n3504}
\begin{tabular}{lll}
\hline
\hline
  & NGC~2903 & NGC~3504 \\
\hline
$\alpha_{2000}$ & 09h32m09.9s & 11h03m11.21s  \\
$\delta_{2000}$ & +21d31m01s & +27d58m21.0s \\
Type$^a$  & SB(s)d  & (R)SAB(s)ab \\
V$_{lsr}$ (\kms) & 560 & 1534 \\
Distance (Mpc) & 8.6$^b$ & 20$^c$ \\
m$_B^a$ & 9.59 & 11.69 \\
PA &  17\degree$^d$  & 149\degree$^c$ \\
Bar PA & 20\degree$^d$ & 143\degree$^c$ \\
L$_{\mbox{FIR}}^a$ (10$^{9} $L$_\odot$) & 8.8 & 14.3 \\
M(\htwo) ($10^9$\msun) & 1.2$^e$ & 1.4$^e$ \\
SFR$^f$ (\msun $\mbox{ yr}^{-1}$) & 1.5 & 2.5\\
\hline
\end{tabular}
\begin{tabular}{l}
$^a$ From NED \\
$^b$ Telesco \& Harper (1980).\\
$^c$ Kenney et al. (1993).\\
$^d$ Sheth et al. (2002). \\
$^e$ This paper. \\
$^f$ Kennicutt (1998)
\end{tabular}
\end{table}

\section{HCN(1-0) emission in NGC3504}

\begin{figure}
\resizebox{9cm}{8cm}{\includegraphics{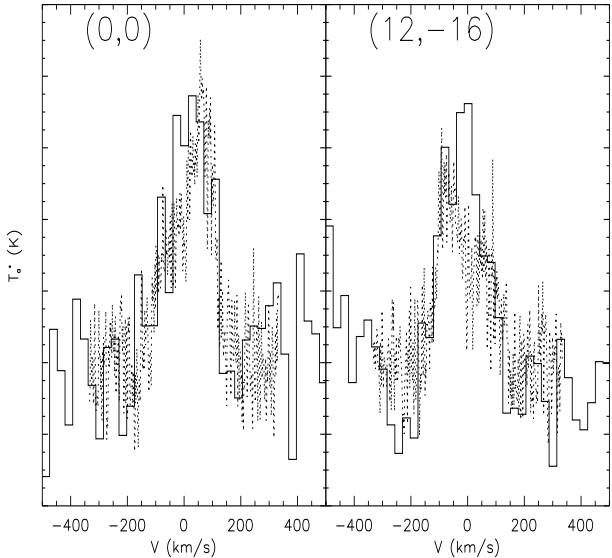}}
\caption{The \hcn  (solid line) and \twelvecotwoone (dotted line) spectra 
 shown for the offset positions (0\arcsec,0\arcsec) and (12\arcsec,-16\arcsec) in the 
galaxy NGC~3504. The \twelvecotwoone intensity is scaled down by a factor of 20 
for comparison.}
\label{fig_n3504_hcn}
\end{figure}

We detected \hcn emission  only at the center (0\arcsec,0\arcsec) 
and  at the offset position (12\arcsec,-16\arcsec), i.e. (20\arcsec,0\arcsec) 
in the bar frame (see Fig. \ref{fig_n3504_hcn}). Here the positive offsets represent
the northern side.
Observed line parameters are listed for these two  positions in Table \ref{tab_n3504_hcn}. 
No \hcn emission was detected 
at  other positions along the bar (PA=143\degree), 
from -60 to 60\arcsec\ in steps of 20 \arcsec\ in the bar frame
and at 10\arcsec, 30\arcsec, 50\arcsec, 70\arcsec\ with $\pm$20 \arcsec\ offset 
perpendicular to the bar, for a typical rms noise temperature of 1.5-2 mK at a 
velocity resolution of 13.6 \kms.
Previous interferometric maps of NGC~3504 by Kohno et al. (1999a)
also reveal a central concentration of \hcn emission,
with a high surface density of molecular gas. 

The ratio of the integrated line intensity of \hcn to that of  the \twelvecoonezero,
$\Re_{\mbox{\small HCN/CO}}$, is an indicator of the molecular gas density and 
temperature (Kohno et al. 1999b).  Using the \twelvecoonezero integrated intensity of
7.55 K \kms\  at the center, detected with  the FCRAO 14m telescope (Young et al., 1995) 
the ratio  $\Re_{\mbox{\small HCN/CO}}$ is estimated to be about 0.12.
However this ratio is an underestimate since the 
FCRAO 14m beam size of (45\arcsec) at the \twelvecoonezero frequency is larger 
than the IRAM-30m beam size (27\arcsec) at the \hcn frequency. Because of the undersampling of the \twelvecotwoone IRAM-30m pointing 
we were not able to constrain the source size using these data.
Indeed with interferometric observations, Kohno et al. (1999a) found a line 
ratio $\Re_{\mbox{\small HCN/CO}}$ of 0.3 in the
center of the galaxy which is consistent with our value if the beam dilution
factor is taken into account (about 2.8)  assuming that the central HCN and CO emission is 
unresolved.


If HCN is tracing the dense phase of the molecular gas,
the mass of the dense molecular gas can be estimated from the \hcn integrated 
intensity. Following Gao \& Solomon  (2004) we use the relation,
\begin{displaymath}
M_{\mbox{HCN}}(\htwo)\approx 10\times L_{\mbox{HCN}} \msun(\mbox{K} \kms \mbox{pc}^2)^{-1}
\end{displaymath}
to estimate the dense molecular gas mass from the HCN luminosity (L$_{\mbox{HCN}}$). 
The HCN luminosity in the area covered by the IRAM beam size  is estimated
to be $2.1 \times 10^7$ K \kms pc$^2$ which gives a molecular mass of 
M$_{\mbox{HCN}}(\htwo) \approx 2.1\times10^8$ \msun. 
The molecular gas mass $M(\htwo)$ is estimated using the standard CO-to-\htwo conversion factor 
(Young et al. 1996),
\begin{displaymath}
M(\htwo) = 1.1\times10^4 D^2 S_{CO} 
\end{displaymath}
where D is the is the distance (Mpc) and $S_{CO}$ is the \twelvecoonezero integrated intensity (Jy.\kms), 
obtained from Young et al. (1995). Thus the molecular gas mass 
in the central 45\arcsec\ is estimated to be about $1.4\times10^9$ \msun.

The dense molecular gas traced by HCN is about 15 \% of the total molecular gas 
in the center of NGC~3504. 
Moreover, high spatial resolution \twelvecoonezero  imaging from 
the IRAM Plateau de Bure Interferometer  indicates that the molecular 
gas is concentrated in the inner 10\arcsec of the galaxy. Thus, NGC~3504 has the high 
concentration of dense molecular gas necessary to feed its central starburst.

\begin{table}
\caption{Observed properties  of \hcn emission in NGC~3504.}
\begin{tabular}{lllll}
\hline
\hline
X offset & Y offset & I(\hcn ) & Velocity & Width \\
(\arcsec ) & (\arcsec ) & K \kms & \kms & \kms \\
\hline
0   &    0  &     1.24  & 19  &  159. \\ 
12  & -16   &    1.12   &  -21 &  149. \\
\hline
\end{tabular}
\label{tab_n3504_hcn}
\end{table}

\section{HCN(1-0) in NGC~2903}

\subsection{HCN(1-0) distribution}

\begin{figure*}
\resizebox{16.5cm}{8.5cm}{\includegraphics{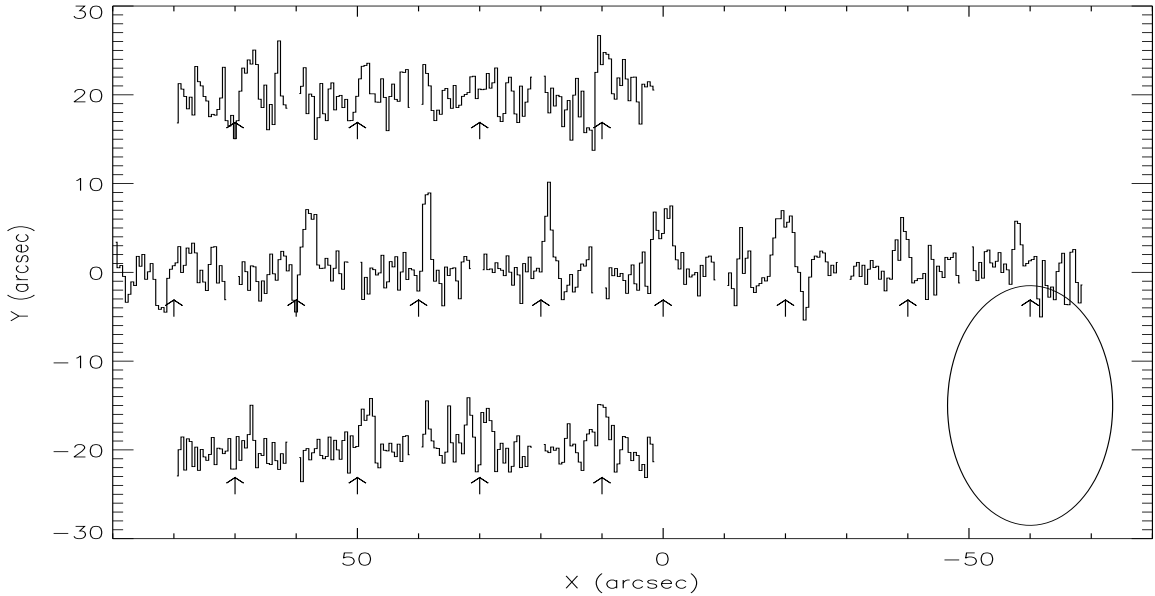}}
\caption{The \hcn spectra along the bar of NGC~2903 
for each pointing. 
The positions are in the frame of the bar. The width of each spectrum ranges from -571 to 567 \kms relative to the systemic velocity of 
the galaxy with a velocity resolution of 27 \kms and the LSR velocity is marked with arrows. The bar axis corresponds to the x-axis. 
The beam size at the \hcn frequency is indicated by the ellipse in the bottom right corner.}
\label{fig_n2903_map_hcn}
\end{figure*}

\hcn emission is detected along the central bar axis of NGC~2903 out to a radius of 60\arcsec.
 The observed intensities for all the pointing positions are given 
in Table \ref{tab_n2903_hcn}.
Toward the center we detect a line intensity of 1.36 K \kms, which is 
about 2.7 times greater than the value obtained by Helfer \& Blitz (1993) 
using the NRAO-12m telescope  with a beam size of 63\arcsec.  This
is likely due to the beam dilution effect. 

\begin{figure}
\resizebox{9cm}{8cm}{\includegraphics{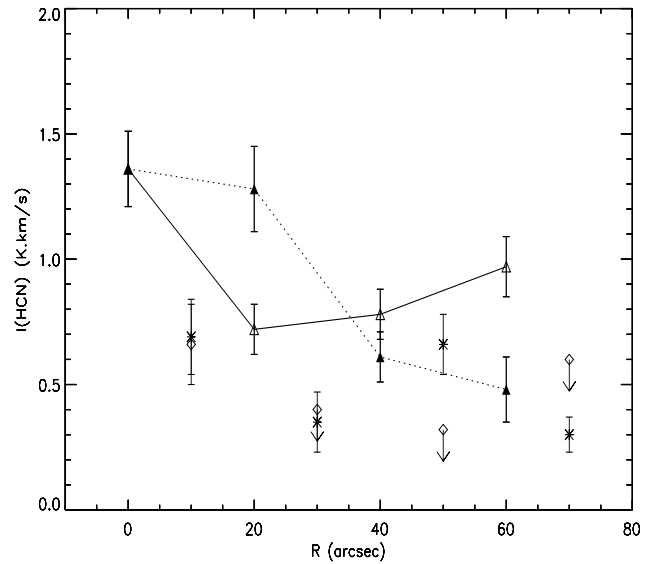}}
\caption{The  integrated intensity of \hcn along the northern side of the bar in NGC~2903 against the offset from the center measured in the frame of the bar. 
The triangles represent the axial positions in the bar (open symbols: northern side, filled symbols: southern side)
, the asterisks 
represent the leading edge positions (-20~\arcsec vertically from the axis of the bar) 
and the diamonds represent the trailing edge positions (20~\arcsec vertically from the 
axis of the bar). The upper limits for the trailing edge positions are indicated by a diamond with a downward arrow, 
taking the velocity width at the same radius in the central part of the bar.}
\label{fig_n2903_hcn_trail_lead}
\end{figure}

On the northern side, the \hcn emission is detected mainly from the central axis 
and the leading edge of the bar (see Fig. \ref{fig_n2903_map_hcn}). 
At a similar rms noise temperature of about 1.5--2 mK, there is no  
detection on the trailing edge at the offsets 
(30\arcsec,20\arcsec),(50\arcsec,20\arcsec) and (70\arcsec,20\arcsec). 
However the only detection of HCN(1–0) at positive offsets from the bar axis. i.e. (10\arcsec,20\arcsec), 
correlates with the CO extension from the center observed by Sheth et al. (2002).
Fig. \ref{fig_n2903_hcn_trail_lead} shows the radial distribution of the 
integrated intensity of the \hcn line  along the bar. 
Besides the strong central emission, there is no radial trend for the \hcn emission 
along the bar. The integrated intensity of \hcn along the
axis of the bar ranging from 0.7 to 1 K \kms is 
slightly higher than that along the leading edge of the bar ranging from  
0.3 to 0.7 K \kms.

The \hcn emission from the leading edge of the bar cannot be  associated with 
the spiral  arm, since the strong spiral arm emerging from the tip of the bar 
is located north of the trailing edge. 

\begin{figure}
\resizebox{9cm}{9cm}{\includegraphics{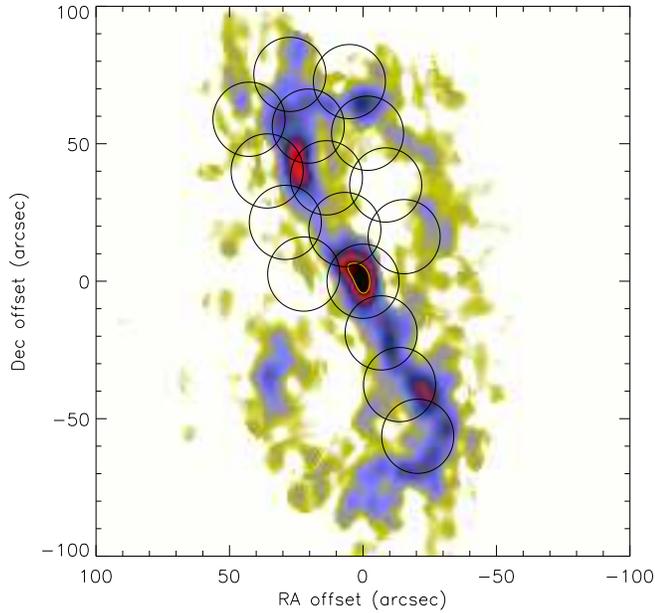}}
\caption{The positions of HCN pointing shown as circles of the size of the beam (27\arcsec) 
is overlaid on the gray scale image of the integrated intensity of 
\twelvecoonezero  emission reproduced from the BIMA SONG survey by Helfer et al. (2003)
}
\label{fig_n2903_cosong_hcnpos}
\end{figure}

In  Fig. \ref{fig_n2903_cosong_hcnpos} , the 12CO(1–0) integrated intensity from the BIMA Survey Of Nearby Galaxies
 is shown together with the \hcn pointings as circles. 
The beam size at the \hcn frequency and the sampling in 
the direction perpendicular to the bar axis does not allow us
to clearly distinguish the origin of the \hcn emission between the central axis and the edges of the bar.
Nevertheless, detection of \hcn on the leading side of the bar indicates
that the \hcn emission is either spatially correlated or downstream of 
the \twelvecoonezero emission.

\subsection{HCN(1-0)/CO(1-0) line ratio}

As a first estimate, the ratio, $\Re_{\mbox{\small HCN/CO}}$, is computed using the \twelvecoonezero integrated intensity of 13.16 K \kms\  
observed towards the center with the FCRAO-14m telescope (Young et al., 1995). 
The ratio $\Re_{\mbox{HCN/CO}}$ at the center is 0.074 which is probably 
an underestimate as in the case of NGC~3504, due to different beam sizes.  
Young et al. (1995) have mapped the bar of NGC~2903 in CO. We used their 
observation at the offset (13\arcsec,43\arcsec) in RA, Dec coordinates and estimated the
ratio, $\Re_{\mbox{HCN/CO}}$, to be about 0.13  at the offset (40,0) in the bar frame. 
This is about twice the size of that at the center.
However, their observations are offset by (0.7\arcsec,5.4\arcsec) 
relative to our pointing and their beam size is about 20\%  larger. 

\begin{figure}
\resizebox{9cm}{8cm}{\includegraphics{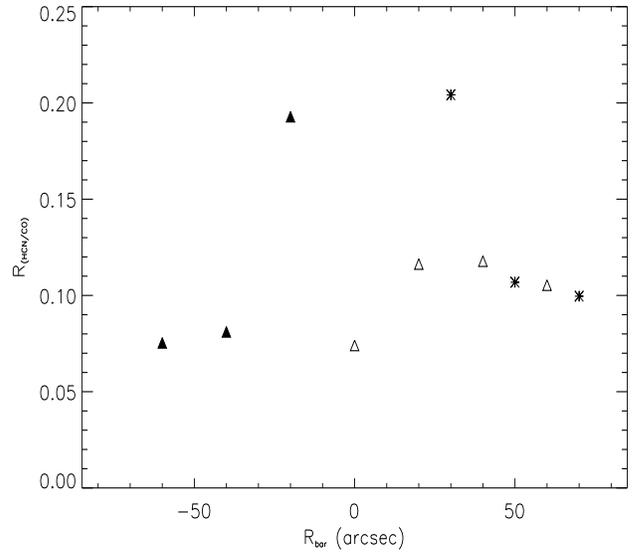}}
\caption{$\Re_{\mbox{HCN/CO}}$ ratio computed using the CO data from 
BIMA-SONG observations (Helfer et al. 2003) and
\hcn from the current observations against the offset measured 
in the frame of the bar.  The symbols have the same meaning as in  
Fig.~\ref{fig_n2903_hcn_trail_lead}
}
\label{fig_ratio_hcn_co}
\end{figure}

Since we have no IRAM-30m single dish \twelvecoonezero observations 
for any of the offsets, we used the interferometric observation
from the BIMA-SONG survey (Helfer et al. 2003) to obtain a more accurate estimate of 
$\Re_{\mbox{\small HCN/CO}}$. These interferometric 
observations have also been combined with single dish observations from the 
NRAO-12m telescope and therefore there is no  missing flux of the large scale 
\twelvecoonezero emission (Sheth et al. 2002). 
The ratio for all the positions are shown in Fig. \ref{fig_ratio_hcn_co}. 
The ratio in the bar ranges from 0.07 to 0.12 without any trend with respect to the 
perpendicular offsets.
Nevertheless a slight drop of the ratio is noticeable along the bar. 
The line ratio on the northern side ranging from 0.09 to 0.12 is larger than 
that on the southern side of $\sim 0.07$. 
This asymmetry in the line ratio between the northern  and the southern sides
of the bar is probably  due to asymmetric physical 
conditions of the gas in the bar. Such asymmetry is seen in 
another barred galaxy, NGC~7479 (Huttemeister et al. 2000), 
where more diffuse gas is present on one side.

On the leading edge of the northern side, at the offset +30\arcsec\ in the bar frame, 
a ratio of $\sim 0.20$ is found. But given the sharp  decrease of CO at 
the edge of the 30m beam at the \hcn frequency, it may be attributed to a pointing offset 
relative to the  interferometric map. 
 Similarly the ratio found at the offset +10\arcsec\ on the 
trailing and leading edges of the bar is very high ($\sim 0.5$) and similarly could be due  to a slight offset
of the \hcn pointings towards the center. They are not displayed in the Fig. \ref{fig_ratio_hcn_co}.

\subsection{Dense molecular gas mass}

The total molecular gas mass in NGC~2903 is estimated with  the BIMA-SONG
\twelvecoonezero emission map as $1.2 \times 10^9$ \msun . All masses based on \twelvecoonezero are calculated using the 
standard conversion factor.
Following a similar procedure as for NGC~3504, we computed the mass of the 
dense molecular gas, M$_{\mbox{HCN}}(\htwo)$, for all the pointings  listed in Table 
\ref{tab_n2903_hcn}.
At the center, M$_{\mbox{HCN}}(\htwo)$ is $2.2\times 10^7$ \msun\ whereas 
the total molecular gas mass is $4.4\times10^8 \msun$ as computed from the 
\twelvecoonezero integrated intensity using the BIMA-SONG  map. 
About 3\% of the molecular gas in the center 
is present in dense clumps. The mass of the dense molecular gas integrated over 
the \hcn pointings along the bar, excluding the central pointing, is $\sim 1.2\times 10^8 \msun$ 
which is about 7.5 \% of 
the total molecular gas mass of $1.6\times 10^9 \msun$ over the same area in the \twelvecoonezero 
BIMA-SONG map. 
It indicates clearly that the formation of dense molecular gas in the bar is more 
efficient than in the center. 

Omitting the southern side of the bar, since we did not observe the 
leading/trailing edge, we computed the dense molecular gas mass and the ratio 
to the total molecular gas mass for the offsets along the bar axis and 
the trailing and leading edges. 
As shown in  Table \ref{tab_mass_hcn} the fraction of dense 
gas is much higher in the leading edge than along the bar axis. 
The value computed for the trailing edge is dominated by the only detection 
at the offset (10\arcsec,20\arcsec) close to the center. For the leading edge, the \hcn pointing
close to the center represents 30 \% of the total gas mass along the leading edge in the bar.
The fraction of dense gas mass along the leading edge 
is two times higher than that along the central axis, within the limitation of the spatial 
resolution of the \hcn observations. 
The bar of NGC~2903 is very different to that of NGC~7479, which is also
classified as a type-B barred galaxy according  to the nomemclature of Martin \& Friedli 
(1997), where Huttemeister et al. (2000) detected \hcn emission only in the center.

\begin{table}
\caption{Observed properties of \hcn emission along the bar of NGC~2903. 
The offset positions X and Y are in the frame of the bar.
The positive X offsets refer to the northern side of the bar. 
The   error ($\sigma$) in I(\hcn) is given between  parentheses.}
\label{tab_n2903_hcn}
\begin{tabular}{llllll}
\hline
\hline
X  & Y  & I$_{\mbox{HCN}}$ & Central velocity & Width & M$_{\mbox{HCN}}(\htwo)$\\
(\arcsec ) & (\arcsec ) & K \kms & \kms & \kms & \msun\\
\hline
0   &    0  &     1.36 (0.15) &      -23. &    200. & 2.2 10$^7$\\  
20  &    0  &      0.72 (0.10) &       -74. &    69.& 1.2 10$^7$ \\   
40  &    0  &     0.78 (0.10)  &     -84  &   66.  & 1.3 10$^7$\\   
60  &    0  &     0.97 (0.12)  &     -134 &   118  & 1.6 10$^7$\\   
80  &    0  &     $<0.40 $    &      &      &  $< 6.6$ $10^6$ \\      
-20 &    0  &     1.28 (0.17) &       19.  &    167  & 2.1 10$^7$ \\   
-40 &    0  &     0.61 (0.10) &       48  &    111  & 1.0 10$^7$\\  
-60 &    0  &     0.48 (0.13) &       121 &    69  &  8.0 10$^6$\\     
10  &    20 &     0.66 (0.16) &       -21 &    109 &  1.1 10$^7$ \\   
10  &    -20 &    0.69 (0.15) &       -4  &    128 & 1.1 10$^7$  \\  
30  &    20  &    $< 0.40$  &             &  &  $< 6.6$ $10^6$ \\     
30  &    -20 &    0.35 (0.12) &       -73  &   79  & 5.8 10$^6$  \\  
50  &    20  &    $< 0.32$           &          &      & $< 5.3$ $10^6$ \\ 
50  &    -20 &    0.66 (0.12) &       -96  &   122   & 1.1 10$^7$ \\ 
70   &   20  &    $<0.60$  &    &     &  $< 10^7$\\   
70   &   -20 &    0.30 (0.07) &        -159  &   40. & 5.0 10$^6$\\    
\hline
\end{tabular}
\end{table}

\begin{table}
\caption{The mass of the molecular gas traced by the \hcn and \twelvecoonezero 
emission are listed in the bar of NGC~2903. The total
mass is estimated by summing all the positions along the axis,
leading and trailing edges seperately, excluding the central position. Between  parentheses we give
the $1\sigma$ error.
\label{tab_mass_hcn}
}
\begin{tabular}{llll}
\hline
\hline
Bar location & M$_{\mbox{HCN}}(\htwo)$ & M$_{\mbox{CO}}(\htwo)$ & 
{\tiny $\frac{\mbox{ M$_{\mbox{HCN}}(\htwo)$}}{\mbox{ M$_{\mbox{CO}}(\htwo)$ }}$ } \\
 &  ($10^7 \msun$) & ($10^7 \msun$) &  \\
\hline
Central axis (North)& 4.1 (0.8)  & 62.1 (0.1)  & 6.6 \% \\
Central axis (South)& 3.9 (0.7) & 49.4 (0.2) & 7.9 \% \\
Leading edge & 3.3 (0.8) & 29.8 (0.2)  & 11.0 \% \\
Trailing edge & 1.1 (0.3) & 20.6 (0.2)  & 5.3 \% \\
\hline
\end{tabular}
\end{table}

\subsection{Dynamics}

\begin{figure}
\resizebox{9cm}{8cm}{\includegraphics{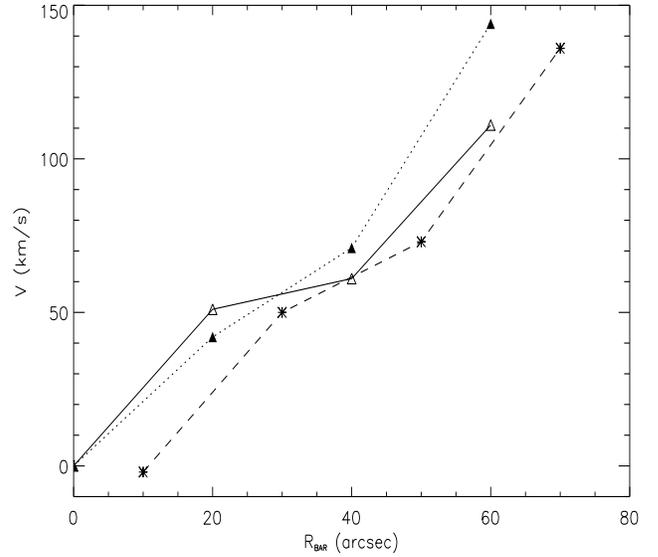}}
\caption{ The modulus of the velocity offset of
\hcn emission along the bar relative to the systemic velocity of NGC~2903
 against the offset measured in the axis of the frame.
The symbols have the same meaning as in  Fig.~\ref{fig_n2903_hcn_trail_lead}. 
}
\label{fig_n2903_hcn_vel}
\end{figure}

\begin{figure}
\rotatebox{270}{\resizebox{8cm}{9cm}{\includegraphics{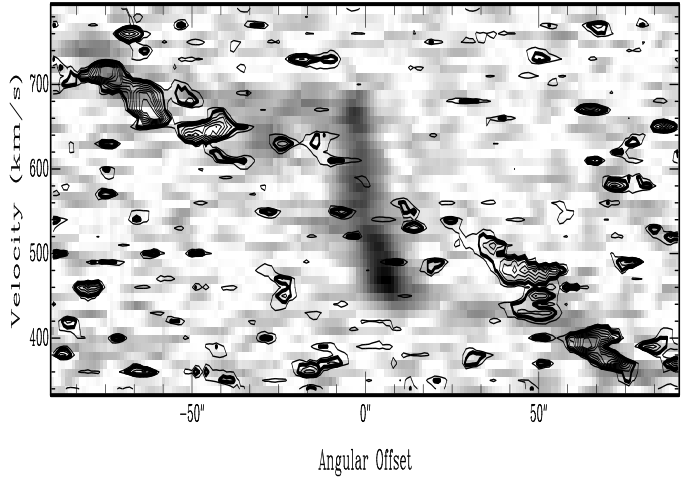}}}
\caption{Position velocity diagram (p-v) of the \twelvecoonezero emission line from the BIMA-SONG map  
along the bar axis (greyscale) and along the northern leading side with positive radii and southern 
trailing side with negative radii (contours).  The offset for the leading/trailing p-v diagram is -20\arcsec
relative to the bar axis. The velocities are in the local standard of rest frame.
}
\label{fig_n2903_song_co_pv}
\end{figure}

Since the bar position angle (20\degree)
is only slightly offset from the galaxy position angle (17\degree), the velocity along the 
bar is a good approximation for the rotation curve, if one neglects any streaming 
motion  along the bar. The velocity differences between the leading and trailing edges (see Fig. \ref{fig_n2903_hcn_vel}) and the \twelvecoonezero position-velocity (p-v)
diagram (see Fig. \ref{fig_n2903_song_co_pv}) from the BIMA-SONG suggest possible streaming motion of 50-80 \kms.
For each offset position of the \hcn emission, the line of sight velocity and 
the velocity width are computed by fitting a Gaussian to the spectra. 

An offset of  23 \kms\ is applied to the observed velocity to symmetrize  the velocities of the receding 
and  approaching sides of the bar. Thus the  heliocentric systemic velocity
of NGC~2903 is  estimated to be 579 \kms.  

The projected rotation curve shown in  Fig. \ref{fig_n2903_hcn_vel} is 
close to symmetric, except at the radius of 60\arcsec\ 
where there is a difference of 33 \kms\ between the approaching and receding side. 
The \hcn velocity in the leading edge of the approaching
side is systematically lower than the bar axis velocity by 5--20 \kms. 
To analyze this velocity offset, the p-v  diagram of the \twelvecoonezero 
line from the BIMA-SONG
map is computed along the bar axis and the northern leading edge (see Fig. \ref{fig_n2903_song_co_pv}). The 
velocity gradient is steeper along the leading edge (positive radii) than the trailing edge which may indicate
that the CO-traced molecular gas is experiencing streaming motion along the leading edge (Athanassoula 1992). 
Indeed the shocks along the bar can create a velocity gradient as large as
100--200 \kms (Athanassoula 1992, Reynaud \& Downes 1997). 
This result is  confirmed by analyzing the p-v diagram on the southern leading edge which exhibits  
an even  steeper gradient on the leading edge.

\begin{figure}
\resizebox{9cm}{8cm}{\includegraphics{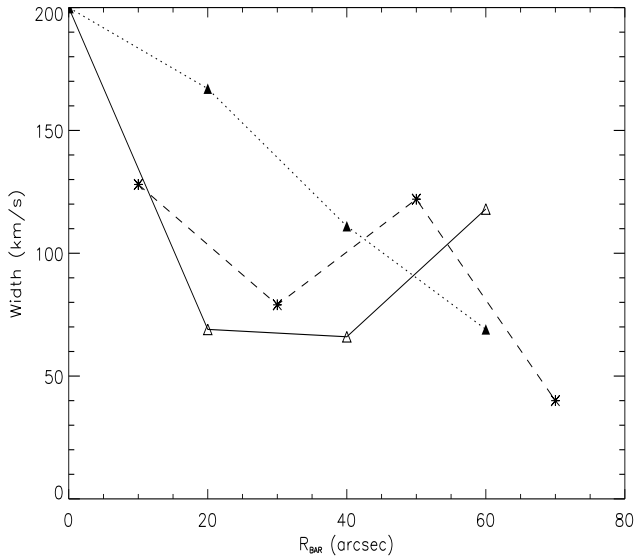}}
\caption{The velocity dispersion of \hcn emission along the bar of NGC~2903 
is shown against the offset measured in the frame of the bar. The symbols
have the same meaning as in Fig.~\ref{fig_n2903_hcn_vel}.
}
\label{fig_n2903_hcn_dispersion}
\end{figure}

The velocity dispersion  of the \hcn spectra 
detected along the bar is shown in Fig. \ref{fig_n2903_hcn_dispersion}. 
The velocity dispersion of  \hcn is about 200 \kms in the center. 
Along the bar the velocity dispersion drops to 40--120\kms, except for the offset (-20,0) 
where the velocity dispersion is 166 \kms. 
The \twelvecoonezero map by Sheth et al. (2002) shows a CO complex slightly south of  
this position (C3 in their nomenclature)  which contributes to the spread in velocity in the \hcn spectra. 

\begin{figure}
\resizebox{9cm}{8cm}{\includegraphics{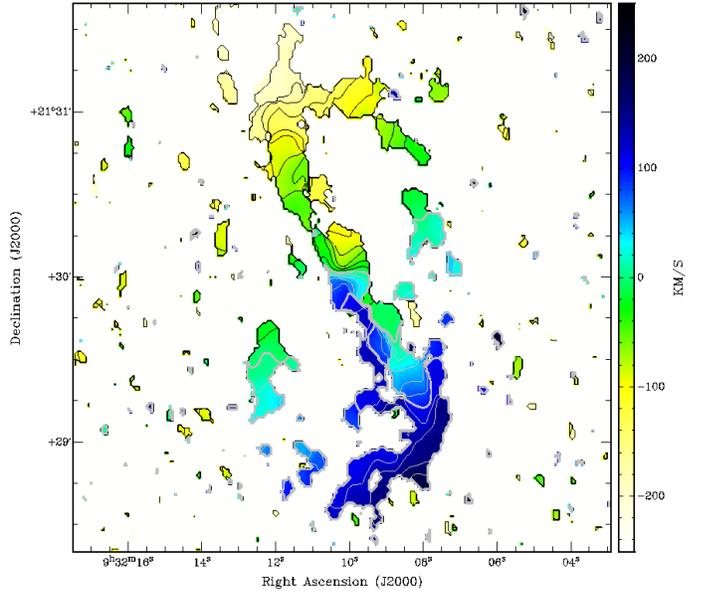}}
\caption{The velocity field of the \twelvecoonezero
emission in NGC~2903 reproduced from the BIMA SONG survey by Helfer et al. (2003)
 shown as contours filled with color scale. T
The spatial resolution is $6.8\arcsec \times 5.5\arcsec$. The levels of
the velocity  are from -200 to 200 \kms in steps of 20 \kms.}
\label{fig_n2903_co_mom1}
\end{figure}

The high velocity dispersion is likely  due to beam smearing of large scale motion can be seen
in the the BIMA-SONG \twelvecoonezero high spatial resolution moment 1 map as shown in 
Fig. \ref{fig_n2903_co_mom1}.
The rotation curve shows an average velocity gradient of about 
2.3 \kms arcsec$^{-1}$, which leads to about 60 \kms across 
the size of the \hcn beam, whereas the typical velocity dispersion for 
GMCs (Sanders et al.  1985) is only a few tens of \kms. This suggests that
the higher velocity dispersion found along the bar is then likely due to 
streaming motion. The \hcn velocity dispersion 
in the leading edge of the bar is slightly higher than that of the bar axis, 
by about 10-20 \kms, except for the largest radius where the velocity dispersion 
is  $\sim 40$ \kms. 
The rotation curve and the velocity dispersion measured using our low resolution
\hcn observations is affected by the velocity gradients due to shock or 
streaming motion. A high spatial resolution observation would be required to understand the role of
shocks in the dynamics of the gas.

\section{Star formation in the bar of NGC 2903}

Comparison of the distribution of \halfa\  emission with the \twelvecoonezero emission
by Sheth et al. (2002) suggests that
the \halfa\ emission is offset further toward the leading
side of the \twelvecoonezero emission, although a bright \halfa\ complex is 
correlated with the peak \twelvecoonezero emission on the northern side of the bar.
On the trailing side of the bar only a few very diffuse \halfa\
regions are detected. But this lack of HII regions cannot be explained by dust 
extinction as indicated by the low molecular gas surface density.

The lack of spatial resolution for the \hcn observations prevents precise comparison 
between the distribution of the \halfa\ emission and the \hcn emission. 
However, detection of \hcn on the leading side of the bar,
and the fact that the \hcn traces the dense molecular gas where SF activity is expected, 
indicate that the \hcn emission could be 
spatially correlated with the HII regions detected by Sheth et al. (2002).

The \halfa\ data, kindly provided by K. Sheth (see Sheth et al. 2002 for more details), 
was calibrated using the scheme of Boselli \& Gavazzi (2002) to obtain a total flux of 
8.5$\times10^{-12}$ \ergcms.  The \halfa\ flux has to be corrected 
for  dust extinction. This correction is rather uncertain since we do not know exactly 
the spatial distribution of dust in the \halfa\ emitting region. 
An estimate of the extinction, A$_V$, from the molecular gas distribution 
(cf. Cernicharo \& Guelin 1987) gives a value higher than 15 in the center. 
Alonso-Herrero et al. (2001) have 
estimated the A$_V$ toward the center ranging from 1 to 6 mag 
using different optical tracers for the HII regions. 
We chose to take an average value of A$_V$=3 mag 
in the center and A$_V$=1 mag in the bar to correct the \halfa\ flux. 
The \halfa\ flux is uncertain within a factor of two 
because of the extinction  and the NII line contribution (Martin \& Friedli 1997).

\begin{figure}
\resizebox{9cm}{9cm}{\includegraphics{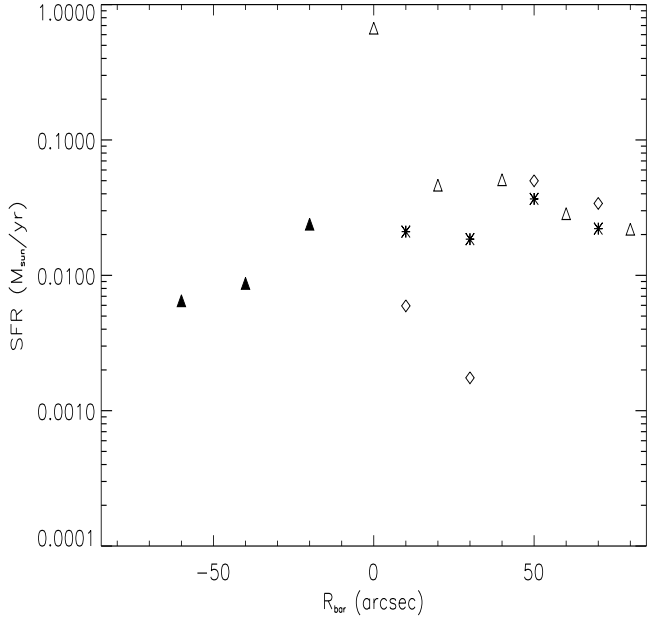}}
\caption{The SFR, in \msunyr, integrated over the IRAM-30m beam at \hcn transition, 
shown against the offset measured in the axis of the frame. 
The symbols have same meaning as in Fig.~\ref{fig_n2903_hcn_trail_lead}.
The SFR is taken above a threshold of 
$1.8\times 10^{-8}$ \msunyr pc$^{-2}$  (3 $\sigma$).}
\label{fig_sfr_pos}
\end{figure}

Using the formula of Kennicutt (1989), the total SFR is estimated to be 
about $\sim$1.4 \msunyr. The SFR estimated over the HCN pointings is 0.36 \msunyr\ in 
the bar and 0.67 \msunyr\ in the center. 
The contrast in SFR between the center and the bar is 1.9. 
The central SFR is consistent with the estimate of Alonso-Herrero et al. (2001) but differs 
in the contrast found between the bar and the center by Knapen et al. (2002) 
because of a different extinction correction. 
We computed the SFR at each HCN offset position within the HCN IRAM-30m beam. 
The local SFR in the center is about 15 to 150 times higher than that of 
the individual pointings in the bar. The SFR in the center is dominated
by the known �hot spots� of recent star formation 
(Oka et al. 1974, Alonso-Herrero et al. 2001). 
The SFR at the HCN pointings in the bar are in the range $\sim 0.01-0.05$ \msunyr.
The SFR  slowly decreases along the bar axis and the leading edge  
as shown in Fig. \ref{fig_sfr_pos}. 
But an inverse trend is present 
for the trailing edge of the bar where the SFR is lower in the inner positions 
(+10 and +30\arcsec) than that in the outer positions (+50 and +70\arcsec). 
These low values at the trailing edge of the bar are likely 
due to a low content of dense molecular gas and a low star formation efficiency. 
For the higher value at the end of the bar,
a contribution from the star formation in the northern spiral arm which falls 
in the HCN beam cannot be ruled out. 

The star formation efficiency can be estimated through the consumption time scale \tausf \
of dense molecular gas,  M$_{\mbox{HCN}}(\htwo)$/SFR. 
In the center \tausf\ is  $\sim 3\times 10^7$ yrs, whereas in the bar \tausf\ 
ranges from 2 to 10 $\times10^8$ yrs for the different positions. 
In the center, the short consumption time and the amount of dense molecular gas 
available for SF does not necessarily imply an earlier starvation of the inner starburst.
The consumption time scale in the bar is much larger than
the dynamical time for the inflow of gas to fuel the center 
(Friedli \& Benz 1995, Martin \& Friedli 1997). 
In addition, a much larger reservoir of molecular gas is available in the center.

\section{Discussion}

\subsection{Ageing of the bar}

Numerical simulations and observations of barred galaxies have shown that the 
SF and the distribution of the gas reveal a wide range of morphologies in the bar 
(Martin \& Friedli 1997, Sheth et al. 2002, Verley et al. 2007).
These different morphologies can be summarized  into three different types,
following Martin \& Friedli (1997): (A) 
the galaxy exhibits SF along the bar without SF activity in the center;
The duty cycle is relatively short, about 500 Myr or less; (B) SF activity is 
present along the bar and in the center of the galaxy; 
Simulations indicate that their duty cycle is  variable
ranging from 300 to 900 Myr. NGC~7479 is a typical example  with intense 
SF along the bar  and in the center (Zurita et al. 2001); 
(C) Very little SF, or none at all, is present along the bar, whereas
intense SF activity is present in the nuclear or circumnuclear region.

The ratio of SFR, $\beta$,  between the center and the bar of NGC~2903 is 1.9.
Comparison with the evolutionary path of $\beta$ in a barred galaxy 
shown in Fig. 4 of Martin \& Friedli (1997) suggests that
NGC~2903 can be classified as a type B galaxy with 
a bar of age between 200 and 600 Myr. The age is dependent on the initial 
conditions in the disk of the galaxy (mass-to-gas ratio, mechanical energy released, SF, bulge mass). 
At the very early stages the models show that intense SF is visible 
along the leading edge of the bar, 
which indicates the presence of dense gas and is indeed observed in our  \hcn data.
Observations of newly born stars along the bar are consistent with the bar being young and the presence of
H$\alpha$ emission in the bar (Sheth et al. 2002) indicates that the bar can be rejuvenated by stars when a new supply 
of gas  becomes available.

When the bar evolves, the SF decreases inside the bar and the $\beta$ increases 
 to typical values of about 20. 
The total lack of \hcn emission along the bar of NGC~3504   indicates that this 
galaxy hosts a much older bar (type C), probably formed more than 1 Gyr ago 
(see Fig. 3 in Martin \& Friedli 1997). 
The dense molecular gas reservoir available in the center of NGC~3504 is about 
ten times higher than that in NGC~2903 which should reach the 
same stage in a few hundred Myr.

\subsection{Dense molecular gas}

Detection of \hcn emission toward the center of galaxies is much more common 
in  surveys of Seyfert galaxies (cf. Kohno et al. 2001, Huttemeister et al. 2000). 
However, very few barred galaxies have been mapped in the \hcn transition
(Reynaud \& Downes 1998; Huttemeister et al. 2000).
We compare the gas properties of NGC~2903 with NGC 7479 and NGC~1530 which
have been classified as type B barred galaxies (Martin \& Friedli 1995). 
In NGC 7479 \hcn emission is seen only in the very center (Huttemeister et al. 2000). 
In NGC~1530, star formation is very intense
around the nucleus and at the ends of the bar, whereas it is weak halfway  between them
but the shocks are more intense (Reynaud \& Downes 1998). 
They concluded that the star formation may be inhibited by the strong shocks and 
the shear. In NGC~2903 the \hcn and the star formation do not show such a 
difference, \hcn is equally strong  along the bar and the star formation does not 
exhibit a clear gradient toward the end, except at the center of the galaxy. 

A striking feature of the current star formation is the strong curve shaped star 
formation region along the bar which indicates that SF may occur in dense molecular gas 
downstream of the shocked regions.

The galaxy, NGC~2903, currently appears to be a unique case of \hcn 
detection along its bar and at the center. The \hcn is distributed 
preferentially toward the leading edge of the bar. 
Most of the HII regions observed  in the bar of NGC~2903 by Sheth et al. (2002) 
are downstream of the CO emitting regions. 
The association of the current star formation with dense molecular 
gas is a further indication that the \hcn is located downstream of the CO regions. 
The large velocity dispersion of the \hcn line indicates that the dense molecular gas
may be located in regions with large velocity gradients, such as shocked regions.
Shocks are expected to trace the dust lane in the bar (Athanassoula 1992).
With the spatial resolution of the \hcn observations presented here, 
the location of the dense gas relative to the shocks cannot be established.

We also find an asymmetry  in the line ratio, $\Re_{\mbox{HCN/CO}}$,  and
the SFR between the northern and the southern sides of the bar. 
The physical conditions and the dynamics of the dense molecular gas 
appear to be complex, and asymmetric along the bar. Higher spatial 
resolution observations of \hcn are necessary  to clarify the situation.

\section{Conclusion}
We have mapped the \hcn emission along the bar of the spiral galaxies 
NGC~2903 and NGC~3504. 

From these observations we can draw the following conclusions:

\begin{itemize}

\item \hcn emission has been detected in the center of NGC~3504 and 
no emission is detected in the bar. The \hcn to \twelvecoonezero line ratio
is 0.12. The mass of the dense molecular gas in the center is 
 $2.1\times10^8$ \msun, which is about 15 \% of the total mass of the molecular gas.

\item In NGC~2903, \hcn emission was detected in the center 
and  along the bar. The \hcn to \twelvecoonezero line ratio along the bar 
is found to vary from 0.07 to 0.12. This ratio is much lower in the center of the galaxy
than in the bar on the northern side, suggesting that the fraction of
dense gas in the bar is higher than that in the center. There is also an asymmetry in
the \hcn emission between the northern and the southern side of the bar.

\item The total dense molecular gas mass in the bar of NGC~2903 is
$\sim 1.2\times 10^8$ \msun\ which is about 6 times larger than the mass
at the center of $\sim 2.2\times10^7$\msun. The \hcn emission is concentrated 
along the central axis and the leading edge of the bar.
The fraction of dense molecular gas mass relative to the total molecular mass is 
 higher along the leading edge  of the bar (11\%) than on the bar axis ($\sim$7\%). 

\item The central velocity of \hcn emission is systematically lower by about 5 to 20 \kms 
in the leading edge of the bar than on the central axis of the bar.
This drop is likely due to streaming motion along the bar. 
A large velocity dispersion of \hcn emission ranging from 40 to 170 \kms\  is
observed along the bar. Comparison with the  \twelvecoonezero rotation curve suggests
that this could be due to beam smearing of the rotational velocity over large scales,
observed with a low spatial resolution.

\item 
Intense SF activity seen in \halfa\  is  correlated with the dense molecular gas. 
The total star formation rate is estimated to be about 1.4 \msunyr, 
with 0.67 \msunyr\ in the center and 0.36 \msunyr\ in the
bar of NGC~2903. The consumption time of the dense molecular gas in the center of about 
3$\times 10^7$ yrs is much shorter than that along the bar, 
ranging from 2 to  10$\times 10^8$ yrs. The dynamical time for the inflow of gas 
from the bar to the center is shorter than the consumption time scale suggesting 
that the inflow of the gas from the bar to
the center will sustain SF activity in the center for a longer period.

\item Comparison of the distribution of \hcn in the bars of NGC~3504 and NGC~2903 
with the results from the numerical simulations of Martin \& Friedli (1997)
suggest that the bars of these galaxies are at different stages of evolution. 
The bar in NGC~3504 is of type C with  age $\ge$1 Gyr, with most of the gas 
already in its center. NGC~2903 harbors a young type B bar with an age 
between 200 and 600 Myr, with a strong inflow of gas toward the center.

\end{itemize}

\begin{acknowledgements}
We thank the anonymous referee for a careful reading and very detailed report which helped to improve this paper significantly
LVM and SL are   partially supported by DGI (Spain) Grant
AYA 2002-03338 and AYA 2005-07516-CO2-02 and  SL was  supported by an Averroes fellowship  
from the Junta de Andaluc\'{\i}a. We thank S. Verley and K. Seth for their 
very useful comments on this work.
This research has made use of the NASA/IPAC Extragalactic Database (NED) which is 
operated by the Jet Propulsion Laboratory, California Institute of Technology, under 
contract with the National Aeronautics and Space Administration.
\end{acknowledgements}

\end{document}